\documentclass[12pt]{iopart}
\usepackage{graphicx,epsfig}
\usepackage{esint}
\usepackage{iopams}
\usepackage{lineno,hyperref}
\hypersetup{colorlinks=true}
\usepackage{graphicx}
\usepackage{wrapfig}
\usepackage{subfigure}
\usepackage{cite}
\usepackage{amstext}
\modulolinenumbers[5]

\renewcommand{\vec}[1]{\ensuremath{\mathbf{#1}}} 
\newcommand{\gv}[1]{\ensuremath{\mbox{\boldmath$ #1 $}}} 
\newcommand{\dd}{\mathrm{d}} 
\newcommand{\ii}{\mathrm{i}} 
 
\newcommand{\eee}[1]{\mathrm{e}^{#1}}

\newcommand{\pd}[2]{\frac{\partial #1}{\partial #2}} 
 
\renewcommand{\phi}{\varphi}

\newcommand{\gtt}{g_{00}}

\hypersetup{
 colorlinks=true,
 linktoc=all,
 linkcolor=red,
 citecolor=blue,
 urlcolor = blue}

\begin{document}

\title{Wave propagation in metamaterials mimicking the topology of a cosmic string}

\author{Isabel Fern\'{a}ndez-N\'{u}\~{n}ez$^{1,2}$ and Oleg Bulashenko$^1$}
\address{$^1$ Departament de F\'{i}sica Qu\`{a}ntica i Astrof\'{i}sica, Facultat de
F\'{i}sica, Universitat de Barcelona, Mart\'{i} i Franqu\`{e}s 1, E-08028
Barcelona, Spain.}
\address{$^2$ Institut de Ci\`{e}ncies del Cosmos (ICCUB), Facultat de
F\'{i}sica, Universitat de Barcelona, Mart\'{i} i Franqu\`{e}s 1, E-08028
Barcelona, Spain.}
\ead{oleg.bulashenko@ub.edu}

\begin{abstract}
We study the interference and diffraction of light when it propagates through a
metamaterial medium mimicking the spacetime of a cosmic string, -- a topological
defect with curvature singularity. 
The phenomenon may look like a gravitational analogue of the Aharonov-Bohm effect, since 
the light propagates in a region where the Riemann tensor vanishes being nonetheless affected by the non-zero curvature confined to the string core.
We carry out the full-wave numerical simulation of the metamaterial medium and 
give the analytical interpretation of the results by use of the asymptotic theory of diffraction, which turns out to be in excellent agreement. 
In particular we show that the main features of wave propagation in a medium with conical singularity can be explained by four-wave interference involving two geometrical-optics and two diffracted waves.
\end{abstract}

%
\vspace{2pc}


\section{Introduction}

Transformation optics \cite{chen16,burokur16} has become a subject of
considerable interest in the recent years given that it provides the ability to
control electromagnetic waves in an unprecedented way. 
The analogy between media and geometry, combined with the advances in the
design of structured metamaterials \cite{tong18}, has motivated extensive
interest for researchers to
develop novel optical devices with unusual properties, some of them inspired by
cosmological models, such as omnidirectional light concentrators
\cite{narimanov09, genov09, cheng10, yin13, sheng13,prokop16},
optical black holes \cite{chen10,dehdashti15,pla-bhole16,bekenstein17},
Hawking radiation in the laboratory \cite{leonhardt15},
optical wormholes \cite{greenleaf07},
optical cosmic strings \cite{zhang15,shi15}, among others.
On the other hand, topological defects
are of special interest for photonics.
They can strongly influence the light-matter interaction, may induce
singularities in the light fields, generate optical beams with orbital angular
momenta, produce matter vortices, etc.\ (see \cite{soskin17,lloyd17} and
references therein).

In this paper, we address the problem of wave propagation in a metamaterial
mimicking the geometry of a topological defect with conical curvature.
On the one hand, our study is largely motivated by the corresponding
cosmological analogue, -- the cosmic string, -- a topological defect that may have
been formed during a phase transition in the early universe
\cite{kibble76,vilenkin81,vilenkin-shellard94}.
On the other hand, such defects are known to appear in elastic solids and nematic liquid crystals 
(called disclinations or wedge dislocations) and they are actively studied
\cite{katanaev05,kleman08,pereira13,fumeron15,seffen16,fumeron17}.
Recently, it was also pointed out that such media with conical singularities
may find important applications in photonics as omnidirectional beam steering
elements \cite{zhang15}.
The study of this kind of defects in the geometrical-optics limit, or equivalently null geodesics in general relativity, 
has been performed by different approaches given that analytical expressions can be easily obtained \cite{galtsov89,depadua98,zhang15,fumeron15}.
However, the wave picture is more complex and not many detailed studies are available \cite{linet86,suyama06,pla-string16,pla-fresnel17}.
Here, we give the full interference and diffraction description of wave
propagation in a medium with conical curvature, by providing the full-wave
numerical simulation and comparing the results with the analytical theory
developed recently \cite{pla-string16,pla-fresnel17}.
Furthermore, we examine different asymptotic approximations -- the uniform
asymptotic theory \cite{boersma68,borovikov} and the geometrical theory of 
diffraction \cite{keller62}, -- applied for this kind of media with conical singularities.
Finally, we study the effects of the wavefront curvature on the results by
contrasting the case of the finite-distance source with that for the plane-wave
incidence.


\section{Cosmic string in a metamaterial medium}\label{sec:meta}
\subsection{Medium parameters}
The development of transformation optics in recent years has established the
relationship between geometry and material properties \cite{chen16,burokur16}.
In particular, it has been shown that light propagation in a curved vacuum
manifold
is formally equivalent to light propagation in a medium filled with an
inhomogeneous anisotropic material embedded in a flat Minkowski spacetime
\cite{leonhardt09}.

To mark out the framework of the model in which we will work, let us consider
the spacetime metric in arbitrary coordinates
\begin{equation}
\dd s^2 = \gtt\dd t^2 + 2g_{0i}\dd t\dd x^i + g_{ij}\dd x^i\dd x^j,
\label{g}
\end{equation}
where $g_{ij}$ is the spatial part of the metric tensor with $i, j = 1, 2, 3$.
Given that our goal is to study wave effects in a spatially anisotropic material,
for simplicity, we consider the case when space and time components of the
metric are decoupled, $g_{0i}=0$.
(For the conical geometry we will consider later on, this assumption corresponds
to the non-spinning cosmic string.)
In this case, there is no coupling between the electric and magnetic fields for the equivalent medium \cite{leonhardt09} and the constitutive relations take the usual form for an anisotropic material
\begin{equation}
D^i=\varepsilon^{ij}E_j, \quad
B^i=\mu^{ij}H_j,
\label{DBEH}
\end{equation}
with the permittivity and permeability tensors given in Cartesian coordinates by
\cite{plebanski60}
\begin{equation}
\varepsilon^{ij}=\mu^{ij}=-\frac{\sqrt{-g}}{g_{00}}g^{ij}.
\label{eq:perm}
\end{equation}
Here, $g^{ij}$ is the inverse of $g_{ij}$, and $g$ is the determinant of the
full spacetime metric.
In this way, the spacetime geometry is mapped into the parameters of a medium.

Next, consider the conical geometry which can be described in cylindrical
coordinates by the line element
\begin{equation}
\dd s^2= -\dd t^2 + \dd r^2 + \beta^2 r^2 \dd \phi^2 + \dd z^2,
\label{eq:metric}
\end{equation}
where $\beta$ is a conical parameter which corresponds to the removal
($0<\beta<1$) or insertion ($\beta>1$) of a wedge with an angle $2\Delta$.
We define $\Delta \equiv \pi (1-\beta)$, given that the removal of the wedge ($\beta<1$) will be the only case studied in this work, therefore $\Delta>0$ is assumed.
The metric \eref{eq:metric} with $\beta=1-4\mu_S$ describes the spacetime of a
static gravitating cosmic string \cite{vilenkin-shellard94} with a mass per unit
length $\mu_S$ confined at the origin $r=0$ along the $z$-axis (by using units
in which $G=c=1$).
It also represents the effective geometry of a linear topological defect in condensed-matter systems (e.g., disclination in nematic liquid crystals \cite{pereira13,fumeron15}, or wedge dislocation in elastic solids \cite{katanaev05}).
It should be noted that cosmic strings formed in the early universe are expected
to have a rather small angular deficit, typically with $\Delta\sim 10^{-7}$
\cite{vilenkin81,vilenkin-shellard94}.
Here, we provide simulation results for the larger scale range, $0<\Delta\lesssim \pi/2$, which is of broader interest in applications and can be easily visualized. 
Yet, our analytical results are valid in the limit $\Delta \ll 1$ as well.

\begin{figure}[h!]
\centering
\includegraphics[width=\textwidth]{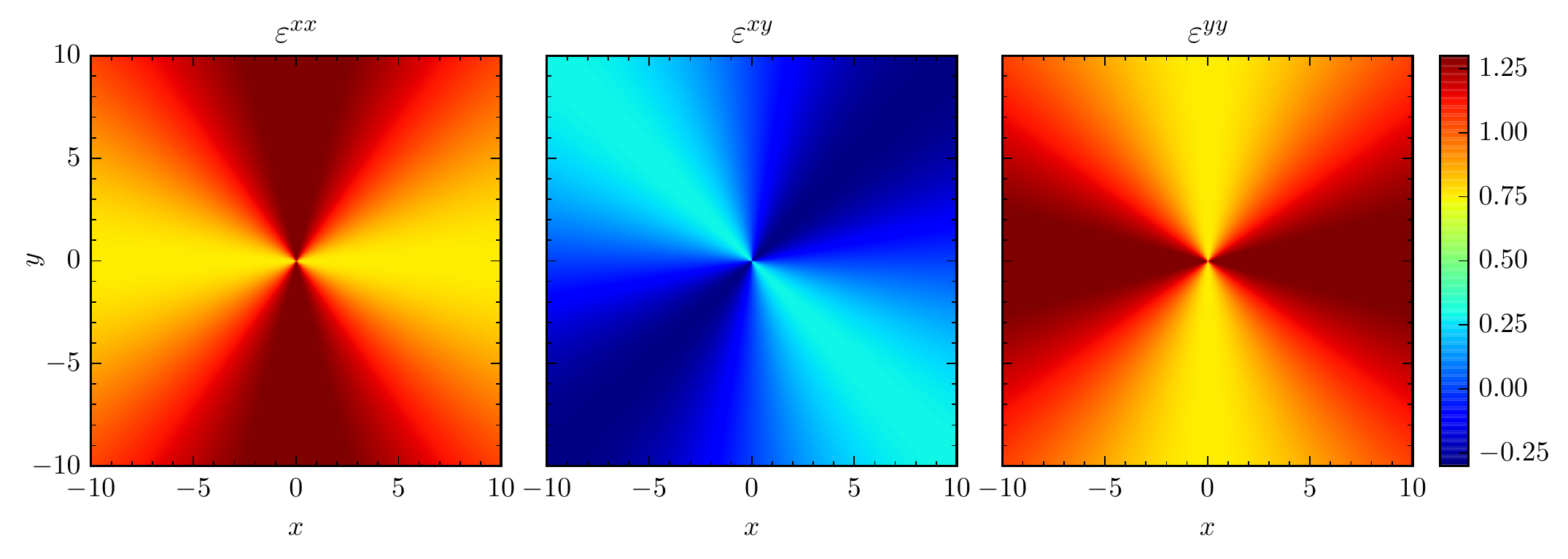}
\caption{Spatial distributions of the components of the medium tensor
$\varepsilon^{ij}$ for the conical geometry with $\beta=3/4$. 
Axes $x$ and $y$ are given in arbitrary units.}\label{fig:param}
\end{figure}

By applying the transformation \eref{eq:perm} to the conical geometry
\eref{eq:metric}, one can find the medium parameter tensors to be diagonal 
in cylindrical coordinates $(r,\phi, z)$, 
\begin{equation}
\varepsilon^{ij}=\mu^{ij}={\rm diag} \left( \beta,1/\beta,\beta \right),
\label{eq:perm-pol}
\end{equation}
which can also be written in the usual Cartesian base $(x,y,z)$ as 

\begin{equation}
\varepsilon^{ij}=\mu^{ij}=\frac{1}{\beta}\left(\begin{array}{ccc}
\beta^2 \cos^2\varphi+ \sin^2\varphi & -(1-\beta^2)\cos\varphi\sin\varphi & 0 \\
-(1-\beta^2)\cos\varphi\sin\varphi & \beta^2 \sin^2\varphi+ \cos^2\varphi & 0 \\
0 & 0 & \beta^2 \end{array}\right).
\label{eq:perm-string}
\end{equation}
Note that the parameters \eref{eq:perm-string} vary with the angle $\varphi$ but are independent of $r$, $z$.
This fact is clearly seen in figure~\ref{fig:param} where the spatial distributions of the tensor are plotted. 
Another distinctive feature is the singularity at the origin, which we will discuss in detail later on.
For the case of a spinning cosmic string, the medium parameters can be found in
Ref.~\cite{mackay10}.

\subsection{Curvature singularity}

A conical spacetime \eref{eq:metric} is an example of a geometry with a
singularity at which the curvature cannot be calculated by standard formulas
\cite{sokolov77,vickers87}.
Yet one can determine the integral of the curvature by applying the Gauss-Bonnet theorem that relates the Gaussian curvature $K$ over some region $\Sigma$ to the geodesic curvature $k_g$ calculated along the smoothly closed boundary $\partial \Sigma$ \cite{mccleary94}:
\begin{equation}
\iint_\Sigma K \, d\sigma = 2\pi - \oint_{\partial \Sigma} k_g \,\dd s.
\label{eq:GB-theorem}
\end{equation}
On the other hand, the rhs of equation \eref{eq:GB-theorem} is the angle $\chi$
a transported frame is rotated through as a result of parallel propagation
around $\partial \Sigma$ \cite{vickers87}:
\begin{equation}
\chi = 2\pi - \oint_{\partial \Sigma} k_g \,\dd s \equiv 2\pi - \tilde{\chi}
\label{eq:paral-tr}
\end{equation}
with $\tilde{\chi}$ being the total geodesic curvature. Intuitively, the
geodesic curvature measures how far a curve is from being a geodesic, the
shortest path between two points \cite{mccleary94}.

\begin{figure}[h!]
\centering
\includegraphics[width=0.65\textwidth]{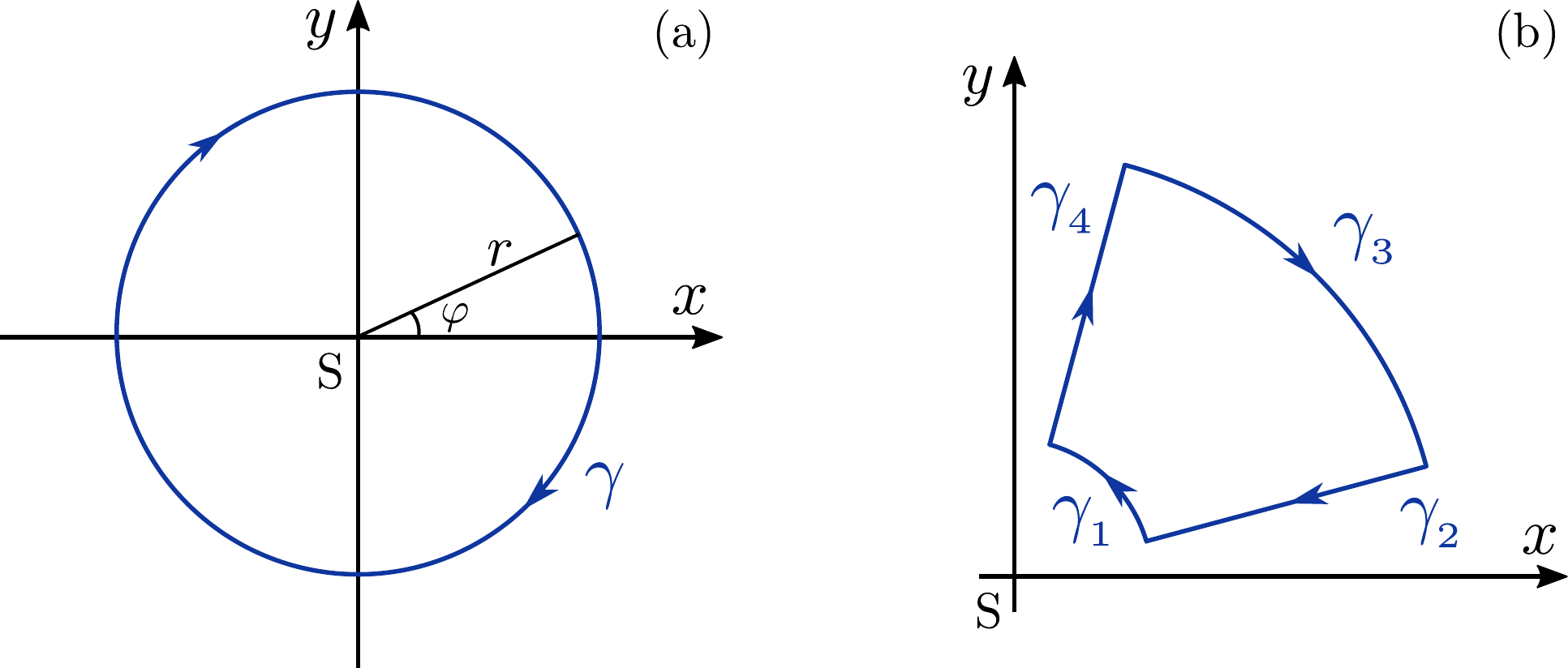}
\caption{Parallel transport along two paths in the spacetime of a cosmic string $S$: (a) around a circle enclosing the singularity at $S$, and (b) outside the singularity.}
\label{fig:loop}
\end{figure}

To calculate the curvature for the metric \eref{eq:metric}, let us determine
$\tilde{\chi}$ over a curve which encloses the singularity.
We take $\Sigma$ to be a disk of radius $R$, at $z=0$ plane, centred at the
singularity $r=0$, and its circumference $\partial \Sigma$ to be a curve
$\gamma$ parametrized by arc-length $s$ [see figure \ref{fig:loop}(a)].
We obtain
\begin{equation}
\tilde{\chi} =
\oint_{\gamma} \Gamma^r_{\varphi\varphi} \,\dot{\varphi}^2 \,\dd s =
\int_0^{2\pi} \Gamma^r_{\varphi\varphi} \,\dot{\varphi} \,\dd \varphi,
\label{eq:geod-curv}
\end{equation}
with $\Gamma^r_{\varphi\varphi}$ being the relevant Christoffel symbol and
$\dot{\varphi} \equiv \dd \varphi / \dd s$ is the angular velocity in units of
the arc-length.
For the metric \eref{eq:metric}, $\Gamma^r_{\varphi\varphi} = -\beta^2 R$ and
$\dot{\varphi}=-1/(\beta R) $, so that the product $\Gamma^r_{\varphi\varphi}
\,\dot{\varphi} = \beta $ is constant. That gives $\tilde{\chi} = 2\pi \beta$
and the frame rotation 
\begin{equation}
\chi = 2\pi (1 - \beta)
\label{eq:chi}
\end{equation}
which is also equal to the angular deficit  $2\Delta$.
Hence, the integral curvature of a disk centred at the singularity is
$2\pi (1 - \beta)$ independently of its radius $R$.

Next, we consider the closed path that does not enclose the singularity, as
shown in figure~\ref{fig:loop}(b).
Since the product $\Gamma^r_{\varphi\varphi}\,\dot{\varphi} = \beta $
calculated on the arc is independent of the radius, the integrals over
$\gamma_1$ and $\gamma_3$, being in the opposite directions, compensate each
other. The paths $\gamma_2$ and $\gamma_4$ are geodesics, therefore they do not
contribute to the geodesic curvature. As a result, for the total path we get
$\tilde{\chi} = 0$.
To apply the Gauss-Bonnet theorem for this case, we have to take into account
that, instead of a smoothly closed curve, we have a piecewise curvilinear
polygon. The theorem still holds, but we must include a correction for the
vertices of the polygon where the curve is not smooth \cite{mccleary94}.
We obtain $\chi= 2\pi -4\cdot (\pi/2)=0$.

Summarizing the above arguments for the two loops, one can conclude that the
Gaussian curvature $K$ for the conical spacetime \eref{eq:metric} can be defined in terms of
the $\delta$-function with its singularity at the origin.
Moreover, in a 2-dimensional manifold, only one component of the Riemannian tensor is independent and can be related to the Gaussian curvature as $K={R^{r\varphi}}_{r\varphi}$ \cite{mccleary94}.
Consequently, the corresponding components of the Riemannian and Ricci tensors can be found in the form
\cite{sokolov77,vickers87} 
\begin{equation}
{R^{r\varphi}}_{r\varphi} = {R^r}_r={R^\varphi}_\varphi= 2\pi \frac{1-\beta}{\beta} \delta_2(r),
\label{eq:riemann}
\end{equation}
where $\delta_2(r)$ is a two-dimensional Dirac $\delta$-function in flat space,
which is equal to $\delta(x)\delta(y)$ in Cartesian coordinates.
In the following, we only consider the case $\beta<1$ that corresponds to a
singularity with positive curvature. The value of $\beta=1$ gives an everywhere
flat Minkowskian spacetime with no curvature.

It should be noted, that the conical singularity is topological in nature and
cannot be eliminated by a coordinate transformation.
Therefore, when we consider a metamaterial analogue in flat spacetime by
applying the transformation \eref{eq:perm} to the conical geometry
\eref{eq:metric}, the singularity should persist, being embedded, in this case,
in the medium parameters.
We also notice an interesting property which follows from the Gauss-Bonnet theorem:
for the conical spacetime no curve enclosing the singularity could be a geodesic line
since it must have a zero geodesic curvature which is in contradiction with equation~\eref{eq:chi}.

\subsection{Wave equation}

Having found the medium parameters which mimic the conical topology of a cosmic
string, we now consider the propagation of a monochromatic electromagnetic wave with
frequency $\omega$ in such a medium.
The Maxwell equations supplemented by the constitutive relations \eref{DBEH}
lead to the wave equation for a time-harmonic electric field vector
$\vec{E}_{\omega}$ in an anisotropic medium \cite{landau-v8}
\begin{equation}
\gv{\nabla} \times
\left[ \underline{\underline{\mu}}^{-1} ( \gv{\nabla} \times \vec{E}_{\omega} )
\right] = \omega^2 \underline{\underline{\varepsilon}} \,\vec{E}_{\omega},
\label{eq:wave}
\end{equation}
where $\underline{\underline{\varepsilon}}$ and
$\underline{\underline{\mu}}^{-1}$ denote the permittivity and inverse
permeability tensors, respectively.
Similar to Ref.~\cite{pla-bhole16}, we consider a transverse electric (TE) wave propagating
at $z=0$ plane,
$\vec{E}_{\omega}(x,y)= E(x,y)\vec{\hat{z}}$,
where $E$ is the out-of-plane $z$ component of the electric field.
Then, the corresponding wave equation in the two-dimensional ($x,y$) space is
given by
\begin{equation}
\pd{}{x}\left(\mu_{xy}\pd{E}{y}-\mu_{yy}\pd{E}{x}\right)-\pd{}{y}\left(\mu_{xx}
\pd{E}{y}-\mu_{xy}\pd{E}{x}\right)=\omega^2\varepsilon^{zz}E,
\label{eq:te}
\end{equation}
where $\mu_{xx}, \mu_{xy}, \mu_{yy}$ are the components of the inverse
permeability tensor $\underline{\underline{\mu}}^{-1}$ and $\varepsilon^{zz}$
corresponds to the permittivity $\underline{\underline{\varepsilon}}$.
A simple inspection of Eq.~\eref{eq:te} shows that the medium parameters can further be simplified.
Indeed, since only one component of $\underline{\underline{\varepsilon}}$ has
entered into the equation, and it is constant [see Eq.~\eref{eq:perm-string}],
the permittivity can be chosen to be isotropic with all its diagonal elements equal to
$\varepsilon^{zz}$ and the off-diagonal terms equal to zero. In this case, only the
magnetic anisotropy is left.
Moreover, one can redefine $\tilde{E}=\beta E$, to obtain
\begin{equation}
\tilde{\varepsilon}^{ij}=\delta^{ij},\quad\tilde{\mu}^{ij}=\delta^{ij}-(1-\beta^2)\left(\begin{array}{ccc}
\cos^2\varphi & \cos\varphi\sin\varphi & 0 \\
\cos\varphi\sin\varphi & \sin^2\varphi & 0 \\
0 & 0 & 1 \end{array}\right),
\label{eq:perm-string2}
\end{equation}
where $\delta^{ij} = {\rm diag} \left( 1,1,1 \right)$.
In this way, the metamaterial has the permittivity $\tilde{\varepsilon}$ of free
space and the permeability $\tilde{\mu}$ inhomogeneous in two dimensions,
simplifying its design.
Finally, if we write the wave equation for this metamaterial in polar
coordinates, we obtain
\begin{equation}
\pd{^2 \tilde{E}}{r^2}+\frac{1}{r}\pd{\tilde{E}}{r}+\frac{1}{\beta^2
r^2}\pd{^2\tilde{E}}{\phi^2}
+\omega^2 \tilde{E}=0,
\label{eq:wave-eqn}
\end{equation}
which is precisely the equation used in
Refs.~\cite{pla-string16,pla-fresnel17,linet86,suyama06} for the cosmic string
background \eref{eq:metric}.

In the next section, we will solve numerically the wave equation \eref{eq:te}
for a TE-polarized wave propagating in a magnetically anisotropic medium.
It should be noted, that the transverse magnetic (TM) wave can also be considered in a similar medium
after the substitutions: $\varepsilon^{ij} \rightleftarrows \mu^{ij}$, $E\to H$.
In such a case, the medium is electrically anisotropic, and one should solve the
equation for the magnetic field $H(x,y)$.


\section{Numerical simulation}

\begin{figure}[t]
\centering
\includegraphics[width=0.4\textwidth]{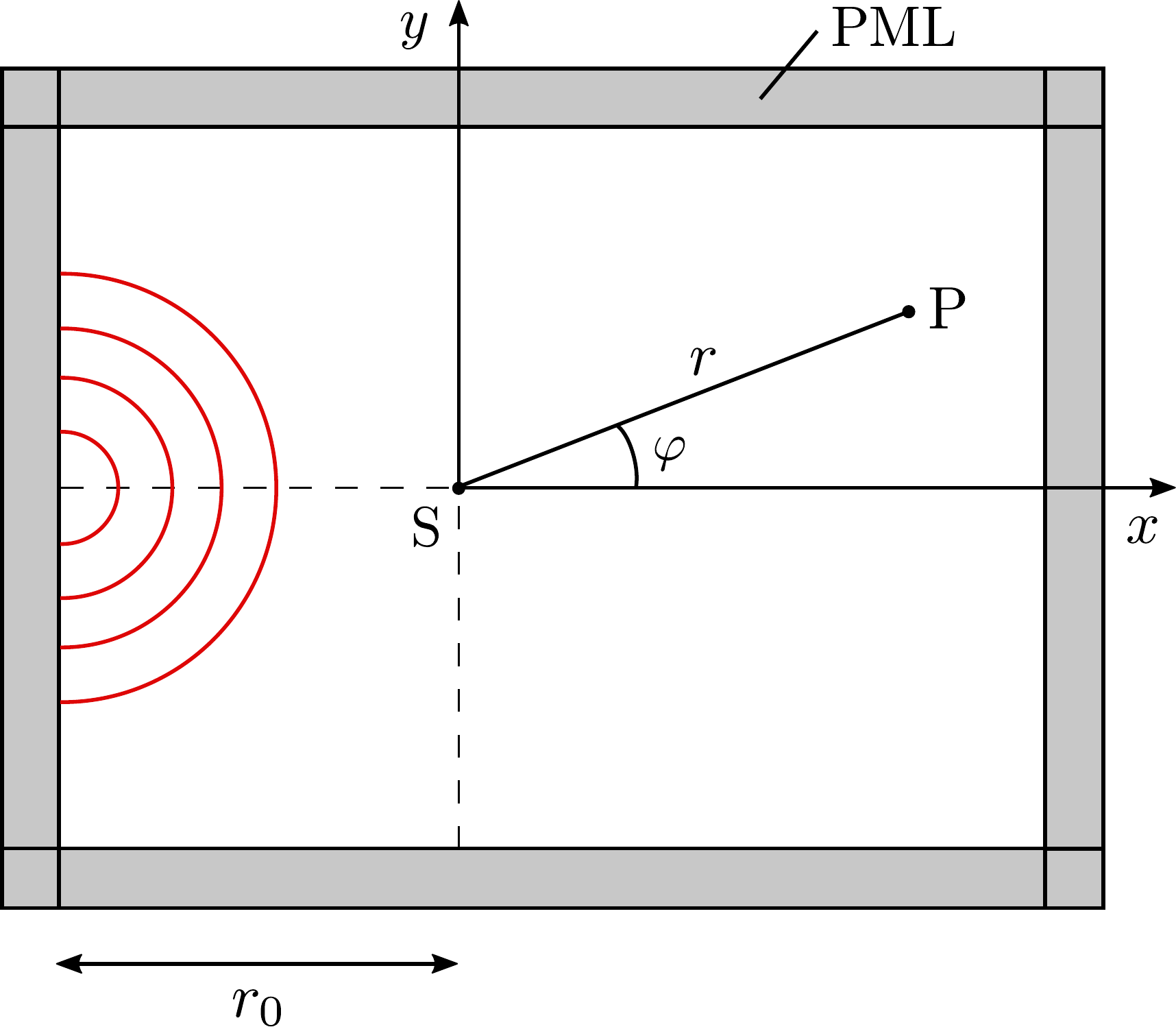}
\caption{Schematic geometry for numerical simulation of wave propagation in a metamaterial mimicking a cosmic string. The wave source is
located at a distance $r_0$ from the string $S$.
The domain is surrounded by a perfectly matched layer (PML) that absorbs the
outward waves to ensure that there are no unwanted reflections. }
\label{fig:geom}
\end{figure}

To solve numerically Eq.~\eref{eq:te} 
we use a rectangular 2D geometry of $(x,y)$ space as shown
in figure \ref{fig:geom}. 
We consider a 2D wave produced by a point source on the left boundary at a
distance $r_0$ from the string $S$.  
The source is an electric line current perpendicular to the domain which
generates TE polarized cylindrical waves.
The wave equation is solved for the electric field $E$.
The distances are normalized by the wavelength $\lambda$, which makes the
equation independent of the frequency. Therefore, it can be solved for any
wavelength as long as it is larger than the unit cells of the metamaterial  and
within its operating bandwidth \cite{burokur16}. 
The same value for the source-string distance $r_0=10\lambda$ is used for all
the simulations in this work.
The simulation was carried out by using the COMSOL Multiphysics 
software package and the figures were processed in Python.
\begin{figure}[b]
\centering
\includegraphics[width=0.9\textwidth]{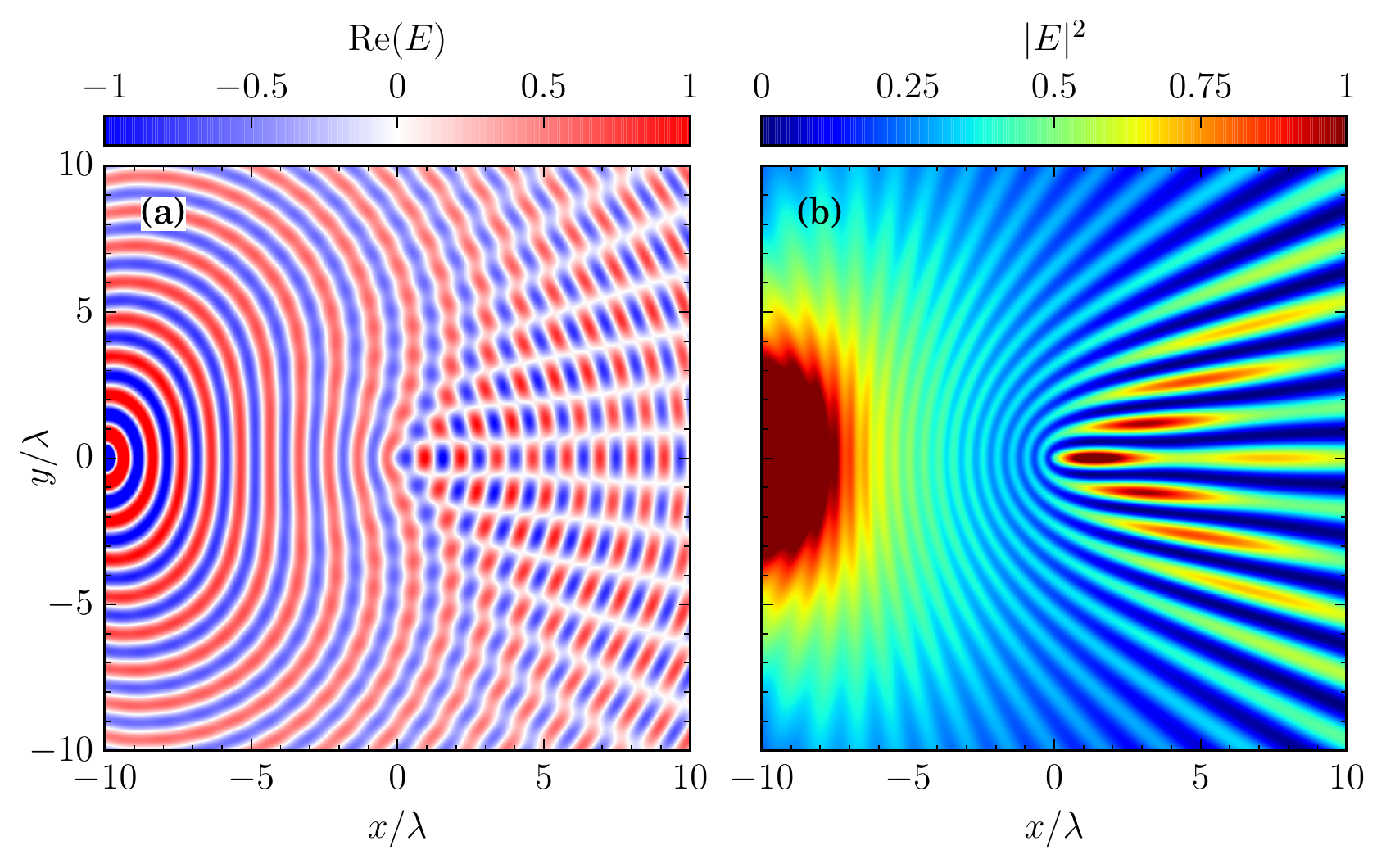}
\caption{
Spatial distributions of the electric field (in arbitrary units) for a medium
mimicking cosmic-string topology:
(a) real part $\mathrm{Re}(E)$; (b) squared norm $|E|^2$.
The wave source is at $(x,y)=(-10\lambda$, 0) and the conical parameters:
$\beta=3/4$, $\Delta=\pi/4$.}
\label{fig:e}
\end{figure}

In figure \ref{fig:e} we show typical results of our simulation for the electric
field $E$ in a medium which mimicks cosmic-string topology.
One can see that the emitted wave propagating around the string interferes with
itself giving rise to a characteristic pattern. This self-interference is
caused by the conical curvature (singularity) at the string location.
By plotting the squared electric field norm in Figure \ref{fig:e}(b), we also
see that the wave field is amplified in some areas behind the string due to
diffraction.
These phenomena will be discussed in detail in the next section along with the
quantitative analysis of the results.
In order to quantify the effect that the string causes on the wave field, we will also
introduce the amplification factor $F\equiv |E/E_0|$, where the electric field $E$ is
normalized by its value $E_0$ in the absence of the string.


\section{Comparison with the analytical theory}
\subsection{Geometrical optics limit}
\begin{figure}[b]
\centering
\includegraphics[width=0.6\textwidth]{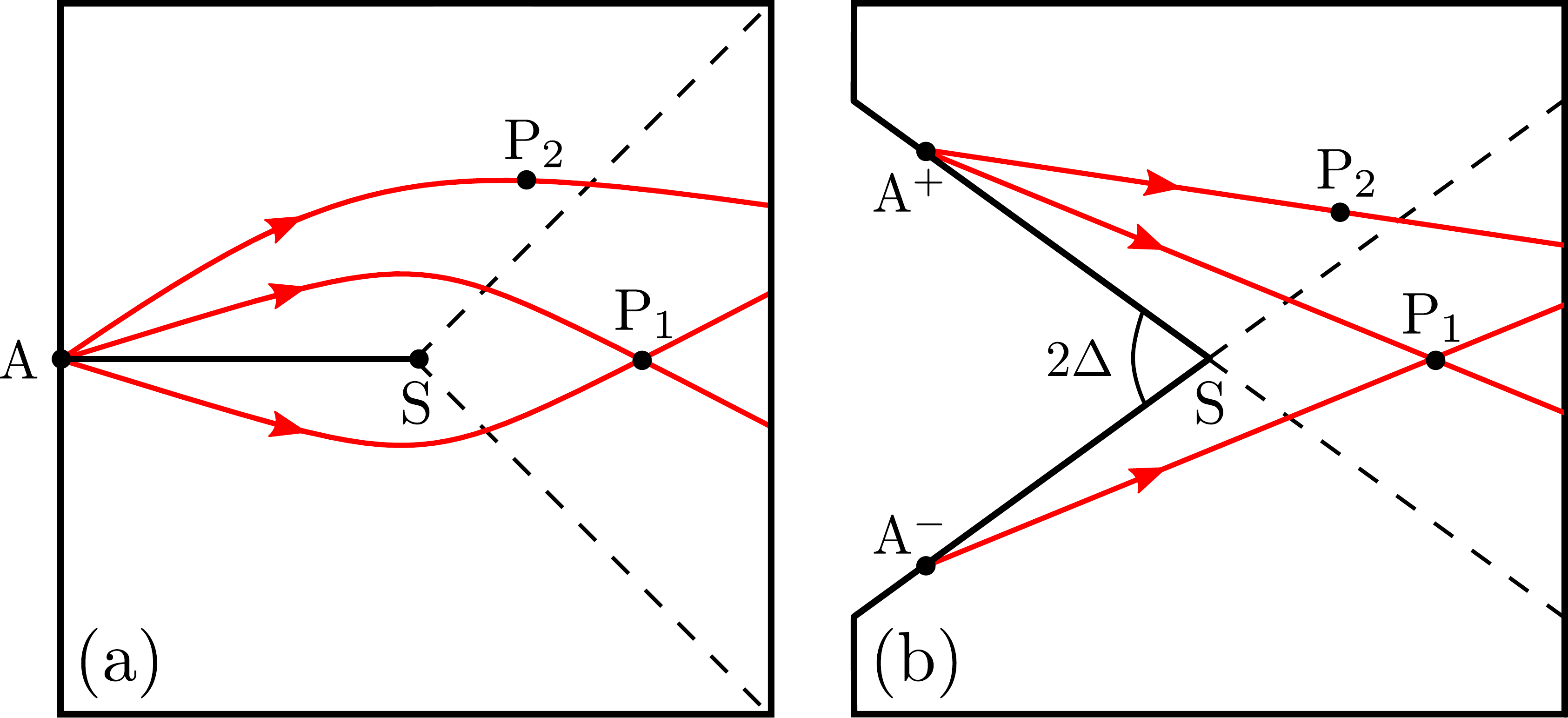}
\caption{Beam traces emitted by a source $A$ in the neighbourhood of the string
$S$ for the geometry: (a) given by the metric \eref{eq:metric} with $\beta<1$; 
(b) Minkowskian geometry ($\beta=1$) with a wedge of $2\Delta$ removed and two
image sources $A^-$, $A^+$ identified. The double-imaging region is bounded by
dashed lines.}
\label{fig:beams}
\end{figure}

First of all, our goal is to understand how the interference pattern observed
in numerics  is formed. We begin with a simple situation: consider a point
source $A$ emitting light beams in the neighbourhood of a string $S$ as shown in
figure~\ref{fig:beams}(a). The conical singularity has the effect of focusing
geodesics whenever $\beta < 1$.
This can be readily seen by changing the angular coordinate to
$\theta=\beta\varphi$, where $\theta$ spans the range $2\pi\beta=2\pi-2\Delta$
in the Minkowski space with a wedge removed and the faces of the wedge
identified \cite{vilenkin-shellard94}.
To perform the angular transformation, we choose to place the cut edge along the
line $AS$ linking the source and the string. In this way, the source $A$ is
doubled to $A^-$ and $A^+$, each on a different side of the wedge, as depicted
in figure \ref{fig:beams}(b) (see \cite{pla-fresnel17} for further details).
In the new representation, the geodesics are simply straight lines and it is
easy to see that the region $-\Delta<\theta<\Delta$, called the double-imaging
region, is illuminated by both image sources $A^-$ and $A^+$. Indeed, an
observer located at $P_1$ in figure \ref{fig:beams} will see two images of the
source $A$. On the other hand, an observer at $P_2$ will only see one image,
being beyond the shadow line for the second one. 
Thus, the conical singularity caused by the string gives rise to a
double-imaging region behind it, in which two image sources interfere.
From the view of figure \ref{fig:beams}(a), there is only one source but the
beams, passing on opposite sides of the string, intersect producing an
interference pattern along and across the line of sight.

\begin{figure}[b]
\centering
\includegraphics[width=0.95\textwidth]{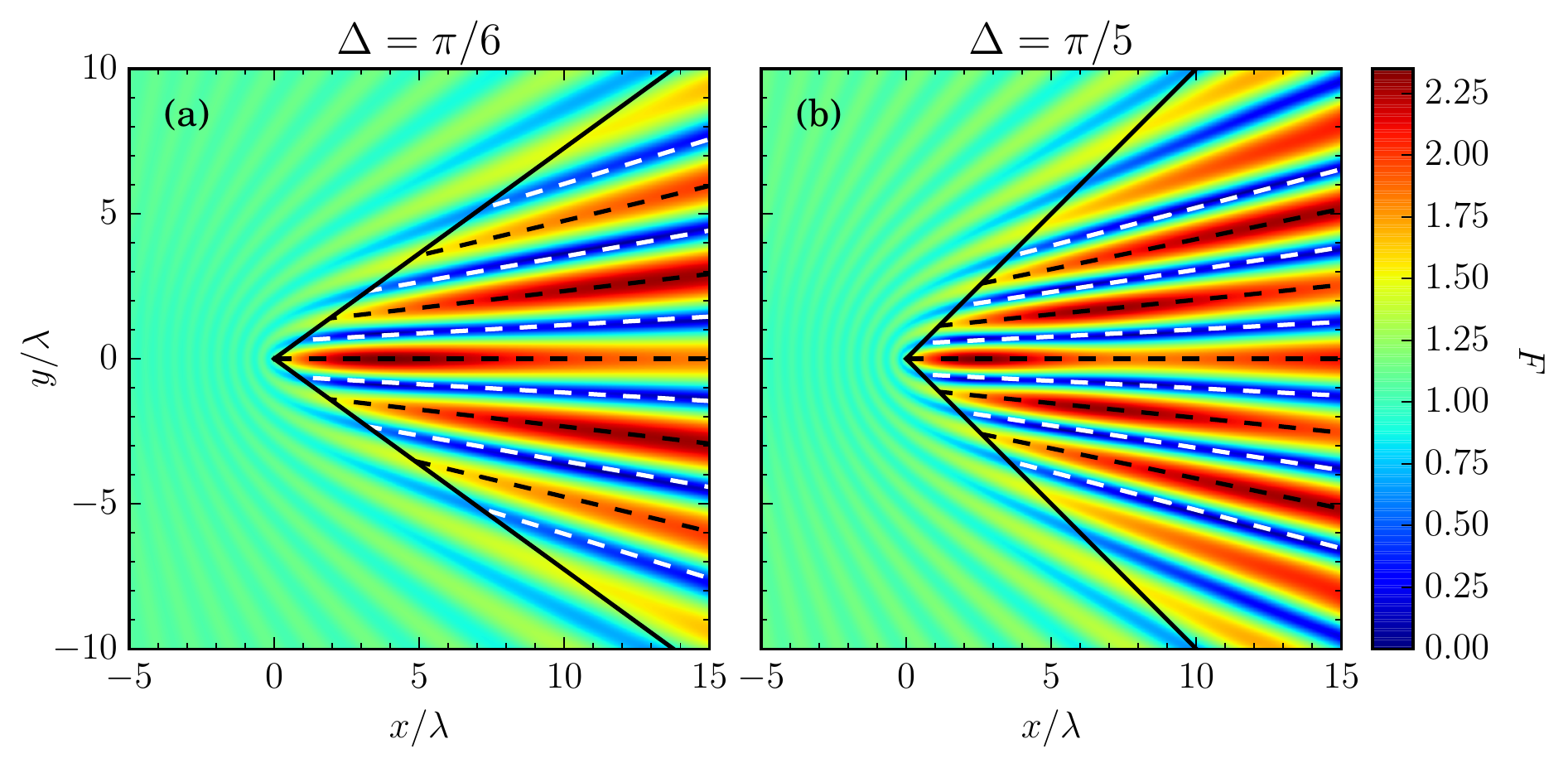}
\caption{Wave pattern for the amplification factor $F=|E/E_0|$ obtained
from numerical simulations in a medium with (a) $\Delta=\pi/6$ and (b)
$\Delta=\pi/5$. 
The nodal and antinodal lines are superimposed by white and black dashed lines,
respectively. Solid black lines are the boundaries of the double-imaging
region.}
\label{fig:nodal}
\end{figure}
The beam interference can be described in the most simple terms as the
geometrical optics (GO) solution. In this limit, the field 
will be the sum of two contributions \cite{pla-fresnel17}
\begin{equation}
E_{GO} = \frac{\eee{\ii ks^-}}{\sqrt{ks^-}}h^-
+ \frac{\eee{\ii ks^+}}{\sqrt{ks^+}}h^+,
\label{eq:go-waves}
\end{equation}
where $k$ is the wavenumber in free space,
$s^\pm=\sqrt{r^2+r_0^2+2rr_0\cos(\Delta\pm\beta\varphi)}$ are the path lengths
of the GO waves from the image sources $A^+$ and $A^-$ to an observation point
[in $(r,\theta)$ space they are cylindrical waves for our geometry], 
and $h^{\pm} \equiv \mathcal{H}(\Delta\pm\beta\varphi)$, with $\mathcal{H}(x)$
being the Heaviside step function that guarantees that the double-imaging range
is determined by $-\Delta/\beta<\varphi<\Delta/\beta$.
Equation \eref{eq:go-waves} allows to understand the appearance of bright and
dark lines behind the string (which we will refer to as ``antinodal'' and
``nodal'' lines, respectively) seen in figure \ref{fig:e}. 
Indeed, to have constructive or destructive interference, disregarding the
slowly varying pre exponential factors, the path length difference, $s^- - s^+$,
must be equal to an integer number of half wavelengths
\begin{equation}
 s^- - s^+ = \frac{\lambda}{2} \, q, \quad \text{with}\; q=
\cases{0, \pm 2, \pm 4, \dots \; \text{in antinodal lines,} \\ 
\pm 1, \pm 3, \pm 5,\dots \; \text{in nodal lines.}}
\label{eq:go-interf}
\end{equation}
In figure \ref{fig:nodal} these lines are superimposed on the wave pattern
(shown in color) simulated by solving the wave equation numerically.
A good agreement is observed within the double-imaging region.
However, one cannot explain the inerference pattern outside that region and the
modulation of the intensity over the antinodal lines within the GO
approximation.

\subsection{Geometrical theory of diffraction}

To advance further, we use the geometrical theory of diffraction (GTD)
\cite{keller62} by considering an additional contribution caused by diffraction.
We write the wave field $E$ at a point $(r,\varphi)$ as the sum of the two GO waves
given by equation \eref{eq:go-waves} and two diffracted (D) waves
\begin{equation}
E = E_{GO} + \frac{\eee{\ii kr_0}}{\sqrt{kr_0}} \frac{\eee{\ii kr}}{\sqrt{kr}}
\, (D^- +D^+ ) ,
\label{eq:GTD}
\end{equation}
where $D^\pm$ are the diffraction coefficients (or directivity functions) given by 
\cite{pla-fresnel17}
\begin{equation}
D^\pm=- \frac{\eee{\ii \pi/4}}{2\sqrt{2\pi}}
  \frac{1}{ \sin{[\frac{1}{2}(\Delta\pm\beta\varphi})] }.
\label{eq:diff-kel-pm}
\end{equation}
The physical interpretaion of the D waves is the following: they are the waves
that go from the source to an observer but hitting the string, following the
shortest path \cite{pla-string16,pla-fresnel17}. The coefficients $D^{\pm}$ give
the amplitude of the D wave as a function of the direction $\varphi$ and they
are singular at $\varphi=\pm\Delta/\beta$, the boundary of the double-imaging
region.

\begin{figure}[b]
\centering
\includegraphics[width=0.9\textwidth]{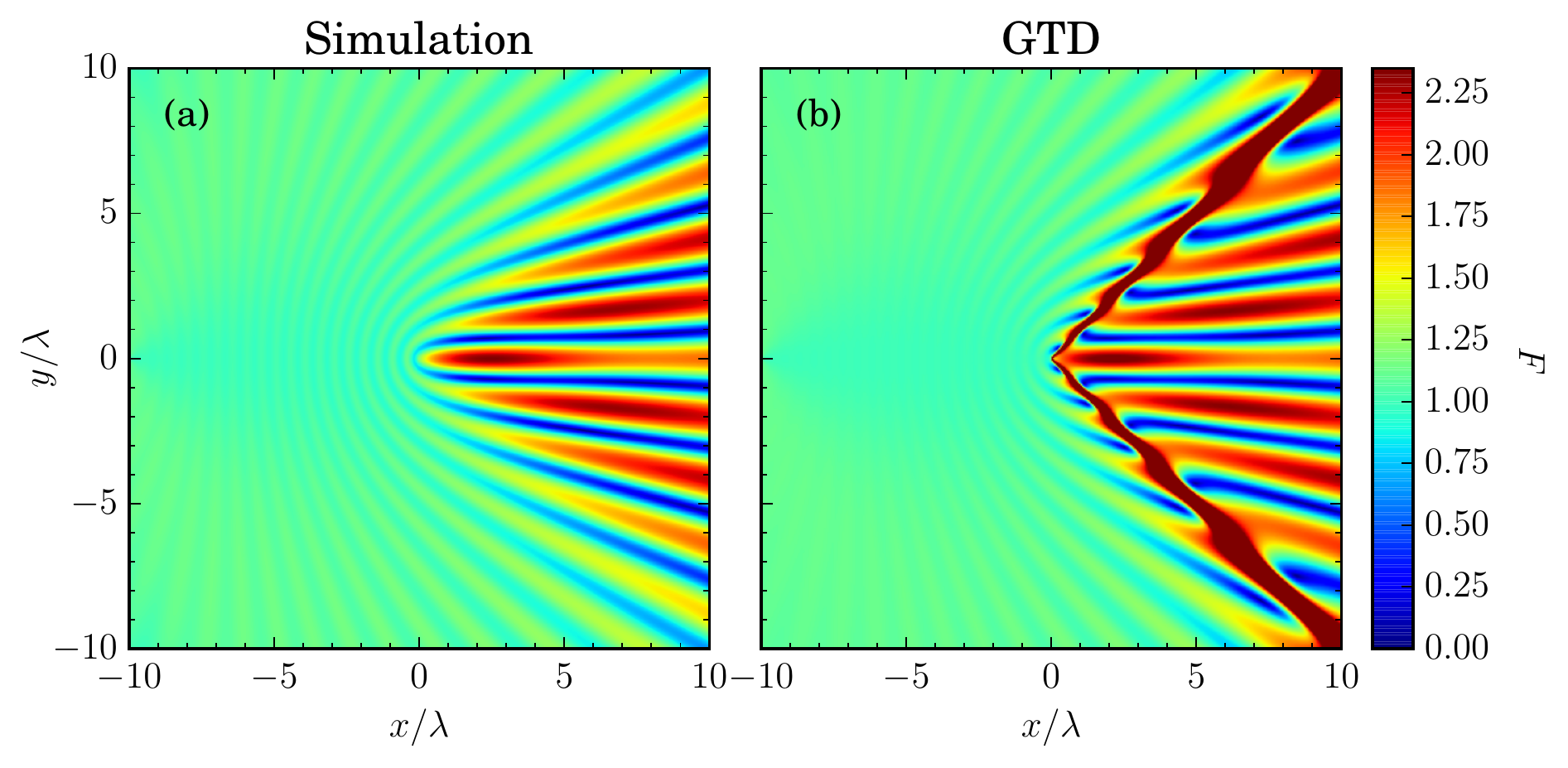}
\caption{Comparison of the wave patterns obtained from
(a) metamaterial simulation, (b) geometrical theory of diffraction (GTD);
for a medium with $\Delta=\pi/5$.}
\label{fig:f-gtd}
\end{figure}
In figure \ref{fig:f-gtd} we compare the results of numerical simulation in the
metamaterial with the GTD solution \eref{eq:GTD}.
It is seen that the wave effects behind the string are very well reproduced
except at the boundary of the double-imaging region, where the D-wave terms
diverge. We also observe that the double-imaging effect results in the field
amplification behind the string and, due to diffraction, the factor $F$ can even
be greater than 2.

It should be emphasized that the D-wave terms are not the next order terms in
the expansion over the wavelength $\lambda$, as it would be in the case of a
diffraction on a compact object. For the conical defect, the relevant parameter is the Fresnel number $N_F\sim r \Delta^2/\lambda$, where $r$ is the distance from the string to the observer and $\Delta$ is proportional to the mass of the string (the singularity strength).
For $\Delta$ and $\lambda$ fixed, one can always find the distance $r$, for
which $N_F \sim 1$, and the D and GO waves are therefore of the same order of
magnitude \cite{pla-fresnel17}. The GO terms dominate whenever $N_F\gg 1$.

The four-wave interference description of the GTD may also be helpful to explain
the intensity modulation over the antinodal lines.
Both D waves follow the path of $r+r_0$, therefore the path difference between
the GO and D waves will be $r+r_0 - s^\pm$.
Also, we have to take into account that the D wave acquires a phase shift of
$3\pi/4$ by hitting the string, as seen in the diffraction coefficients
\eref{eq:diff-kel-pm}. 
\begin{figure}[b]
\centering
\includegraphics[width=0.95\textwidth]{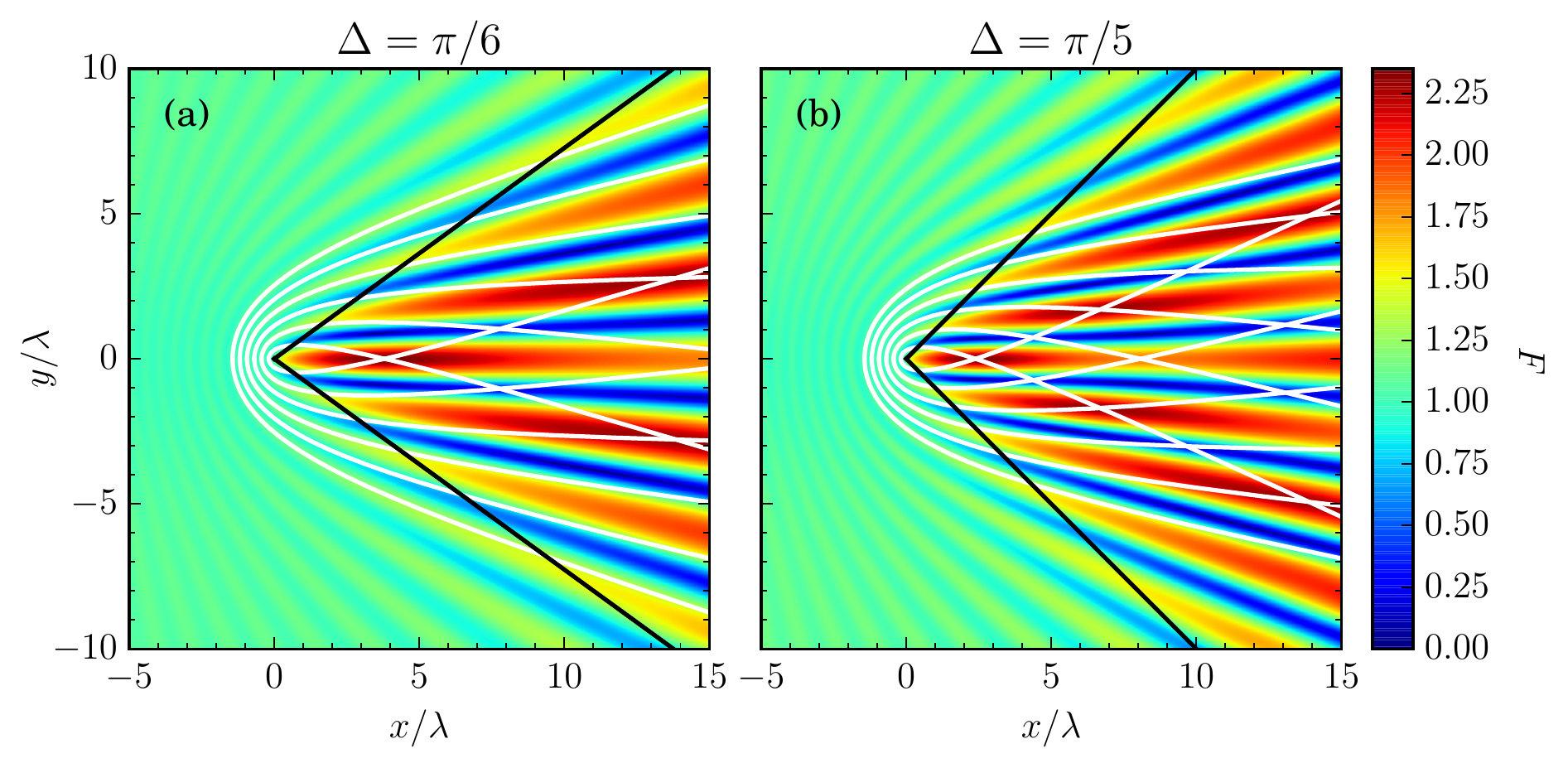}
\caption{Wave pattern for the amplification factor $F$ obtained from numerical
simulation for (a) $\Delta=\pi/6$ and (b) $\Delta=\pi/5$.
The lines of constant phase between the GO and D waves (superimposed by white)
determine the diffraction maxima.}
\label{fig:f-lines}
\end{figure}
Therefore, we would expect the maxima and minima to
occur when these two conditions are fulfilled simultaneously:
\begin{eqnarray}
\label{eq:max-min1}
r+r_0 - s^+  =  \frac{\lambda}{2} \, \left(n+\frac{3}{4} \right), \\
r+r_0 - s^-   =  \frac{\lambda}{2} \, \left(m+\frac{3}{4} \right)
\label{eq:max-min2}
\end{eqnarray}
with $m,n$ being non-negative integers: $0, 1, 2, \dots$. 
These constant-phase lines are hyperbolas in $(r,\theta)$ space 
\cite{pla-fresnel17}.
It is easy to see that, by subtracting equations \eref{eq:max-min1} and
\eref{eq:max-min2}, one obtains equation \eref{eq:go-interf} with $q=n-m$.
That means that the intersection points of the hyperbolas lie on the nodal and
antinodal lines (see figure \ref{fig:f-lines}). The maxima are found whenever
both indices $n$ and $m$ are even, while the saddle points (local minima on the
antinodal lines) correspond to simultaneously odd $n$ and $m$. The interference
points can be labelled with the indices $(n,m)$ and be associated to ``Fresnel
observation zones" (see \cite{pla-fresnel17} for further details). 
Summarizing, the main features of the wave pattern observed in numerics can be
explained as the interference of four characteristic waves of the GTD.
It is known from singular optics \cite{berry02,dennis09} that when three or more
waves interfere in two dimensions, the intensity vanishes at points, rather than
on lines, that may produce optical singularities. This is not the case for the
conical space and the type of waves we consider, where four characteristic waves
(two GO and two D waves) have only two independent phase differences, due to the
symmetry, hence the intensity vanishes on nodal lines.

\subsection{Uniform asymptotic theory}
\begin{figure}[b]
\centering
\includegraphics[width=0.9\textwidth]{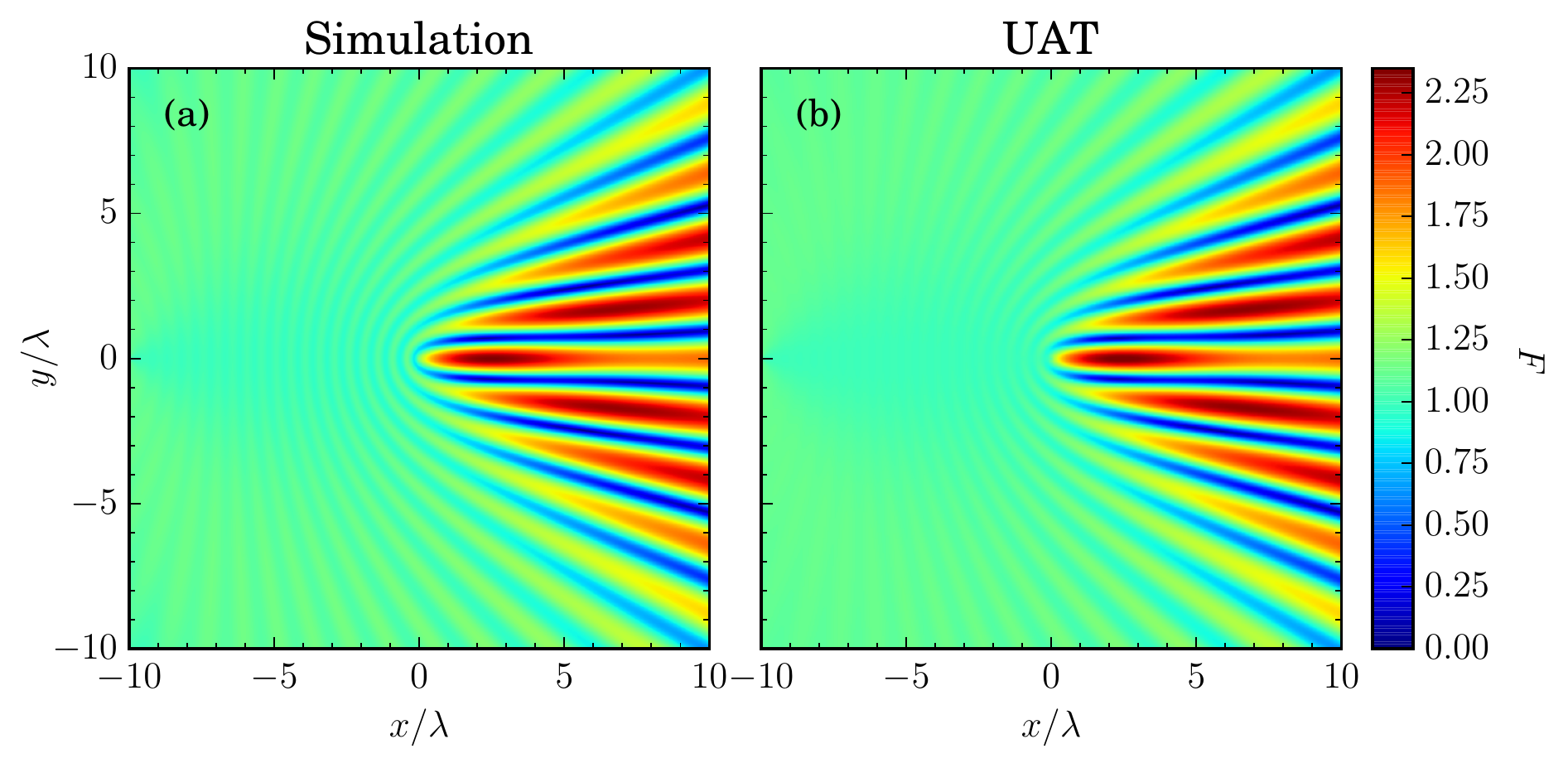}
\caption{Comparison of the wave patterns obtained from: 
(a) metamaterial simulation, (b) uniform asymptotic theory (UAT);
for a medium with $\Delta=\pi/5$.}
\label{fig:f-uat}
\end{figure}
The GTD we have used to explain the numerical results is quite satisfactory 
in the whole region of interest except at the boundary of the double-imaging region where the solution diverges. This nonuniform solution can however be
``regularized'' eliminating the singularity in the framework of the
uniform asymptotic theory (UAT) \cite{boersma68,borovikov}.
We write the solution for the wave field in the form \cite{pla-fresnel17}
\begin{equation}
E = \frac{\eee{\ii ks^-}}{\sqrt{ks^-}}\mathcal{F}(w^-) +
\frac{\eee{\ii ks^+}}{\sqrt{ks^+}}\mathcal{F}(w^+)
+ \frac{\eee{\ii kr_0}}{\sqrt{kr_0}} \frac{\eee{\ii kr}}{\sqrt{kr}}
\left(\tilde{D}^- +\tilde{D}^+ \right) ,
\label{eq:UAT}
\end{equation}
which is the sum of the penumbra field defined in terms of the Fresnel integral
$\mathcal{F}(w) = \eee{-\ii\pi/4}\pi^{-1/2} \int_{-\infty}^w \eee{\ii u^2}\dd
u$ and the diffracted field with modified diffraction coefficients 
\begin{equation}
\tilde{D}^\pm=- \frac{\eee{\ii \pi/4}}{2\sqrt{2\pi}}
 \left[  \frac{1}{ \sin{[\frac{1}{2}(\Delta\pm\beta\varphi})] }
- \sigma^\pm \sqrt{\frac{2rr_0}{s^\pm(r+r_0-s^\pm)}} \right].
\label{eq:diff-pm}
\end{equation}
Here, we use the notations 
$w^\pm=\sigma^\pm\sqrt{k(r+r_0-s^\pm)}$ and 
$\sigma^\pm={\rm sgn}(\Delta\pm\beta\varphi)$.
Note that the solution \eref{eq:UAT} is uniform, since it is finite and
continuous across the boundary of the double-imaging region.
It can be shown  \cite{pla-fresnel17}, that far from the boundary ($w^\pm\gg
1$), both asymptotics, the GTD \eref{eq:GTD}  and UAT \eref{eq:UAT}, coincide.
In figure \ref{fig:f-uat} we compare the results obtained by numerical
simulation in a medium with the UAT solution given by equation \eref{eq:UAT}.
We see an excellent agreement in the entire region.

\begin{figure}[b]
\centering
\includegraphics[width=0.9\textwidth]{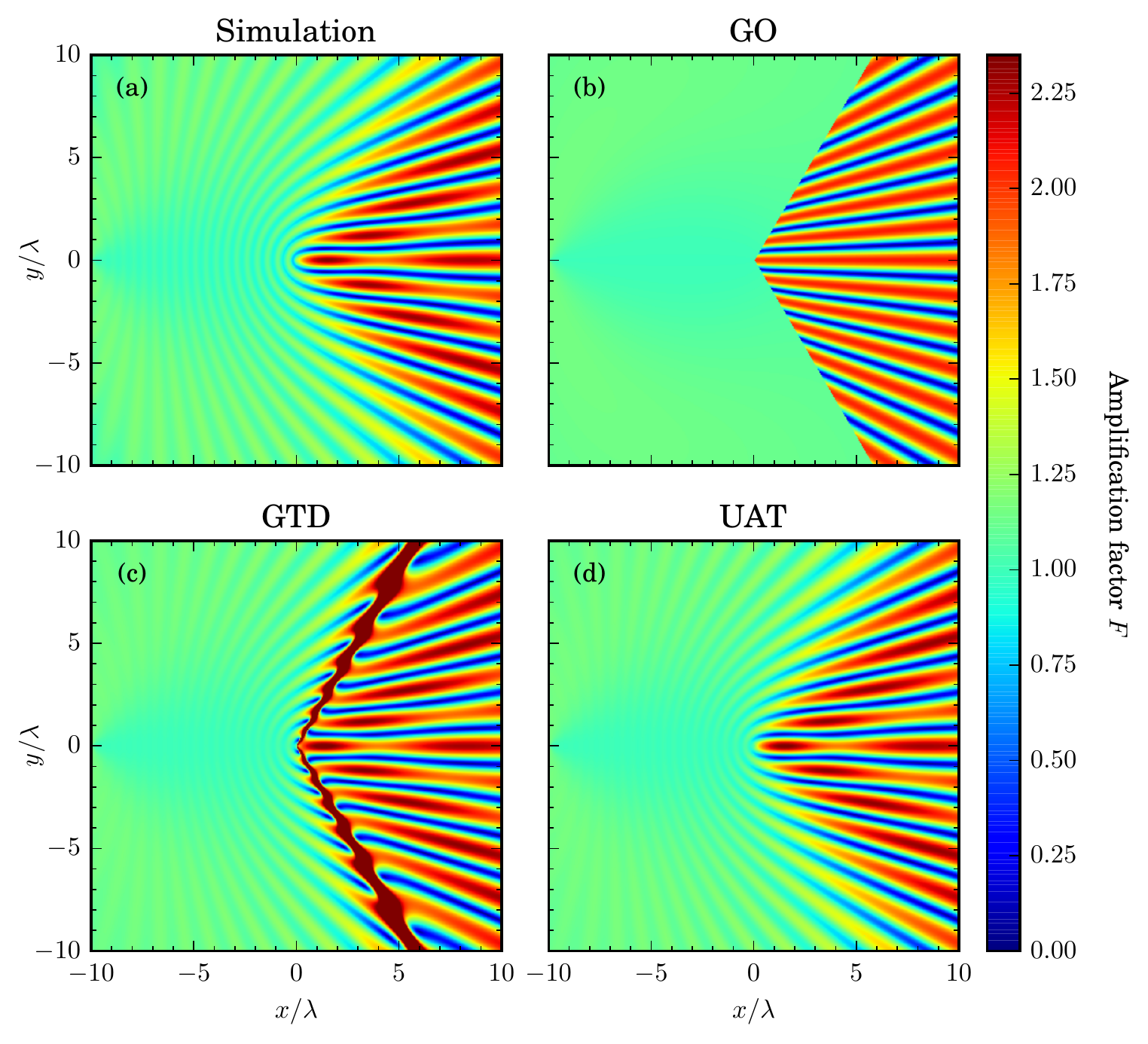}
\caption{Comparison of the wave patterns obtained from: 
(a) metamaterial simulation, (b) geometrical optics, (c) geometrical theory of
diffraction, and (d) uniform asymptotic theory; for a medium with $\Delta=\pi/4$.}
\label{fig:f-all}
\end{figure}
\begin{figure}[h!]
\centering
\includegraphics[width=0.9\textwidth]{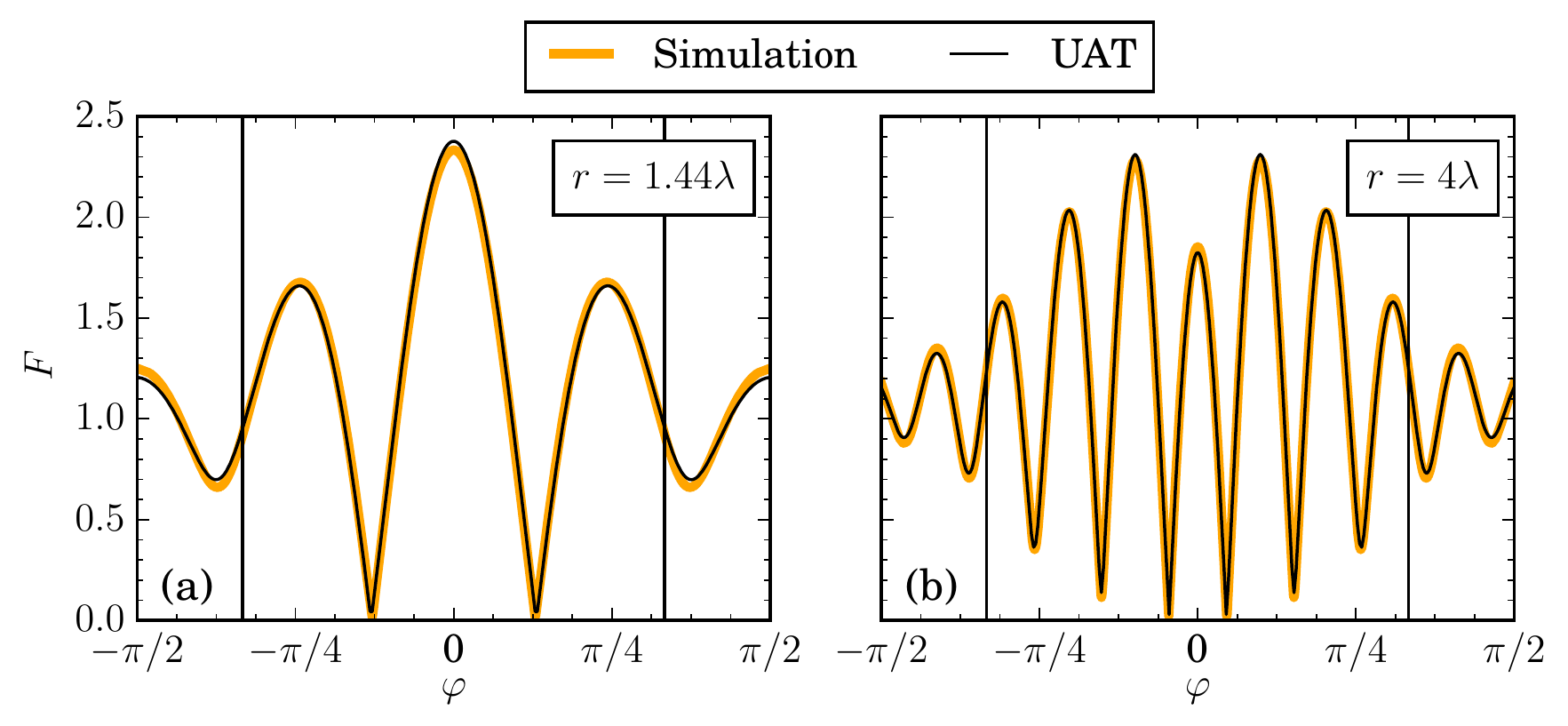}
\caption{Comparison of 1D plots for the factor $F$ of figures \ref{fig:f-all}(a) 
and \ref{fig:f-all}(d). 
$F$ is shown as a function of the angle $\varphi$ for two fixed distances between the string and
the observation point: (a) $r= 1.44\lambda$, corresponding to the highest diffraction 
maximum, and (b) $r=4\lambda$.
Vertical black lines define the boundaries of the double-imaging region.}
\label{fig:f-compare-1D}
\end{figure}

The numerical results for another value of $\Delta=\pi/4$ are shown in figure
\ref{fig:f-all}  in comparison with the analytical theories discussed above.
This figure allows to compare all types of approximations in one view and
observe the corrections that each one gives.
With the GO approximation [figure \ref{fig:f-all}(b)] one can already observe
the main interference effects in the double-imaging region. 
Nevertheless, there is an abrupt change of the field at the boundary between 
the single- and double-imaging zones with no wave effects in the former.
The wave pattern is substantially improved by use of the diffraction terms in
the GTD, which also add the modulation in the antinodal lines [figure
\ref{fig:f-all}(c)]. Lastly, the UAT [figure \ref{fig:f-all}(d)] provides a
smooth transition between the single- and double-imaging regions, having the
least discrepancy with the numerics.
How small the discrepancy is, one can appreciate from 1D plots of figure
\ref{fig:f-compare-1D}, where the amplification $F$ is shown as a function of
$\varphi$ for two fixed distances $r$.
One can see that the agreement between the metamaterial simulation and the UAT
formula \eref{eq:UAT} is not only qualitative but also quantitative.

\section{Plane wave propagation in conical space}

\begin{figure}[h!]
\centering
\includegraphics[width=0.9\textwidth]{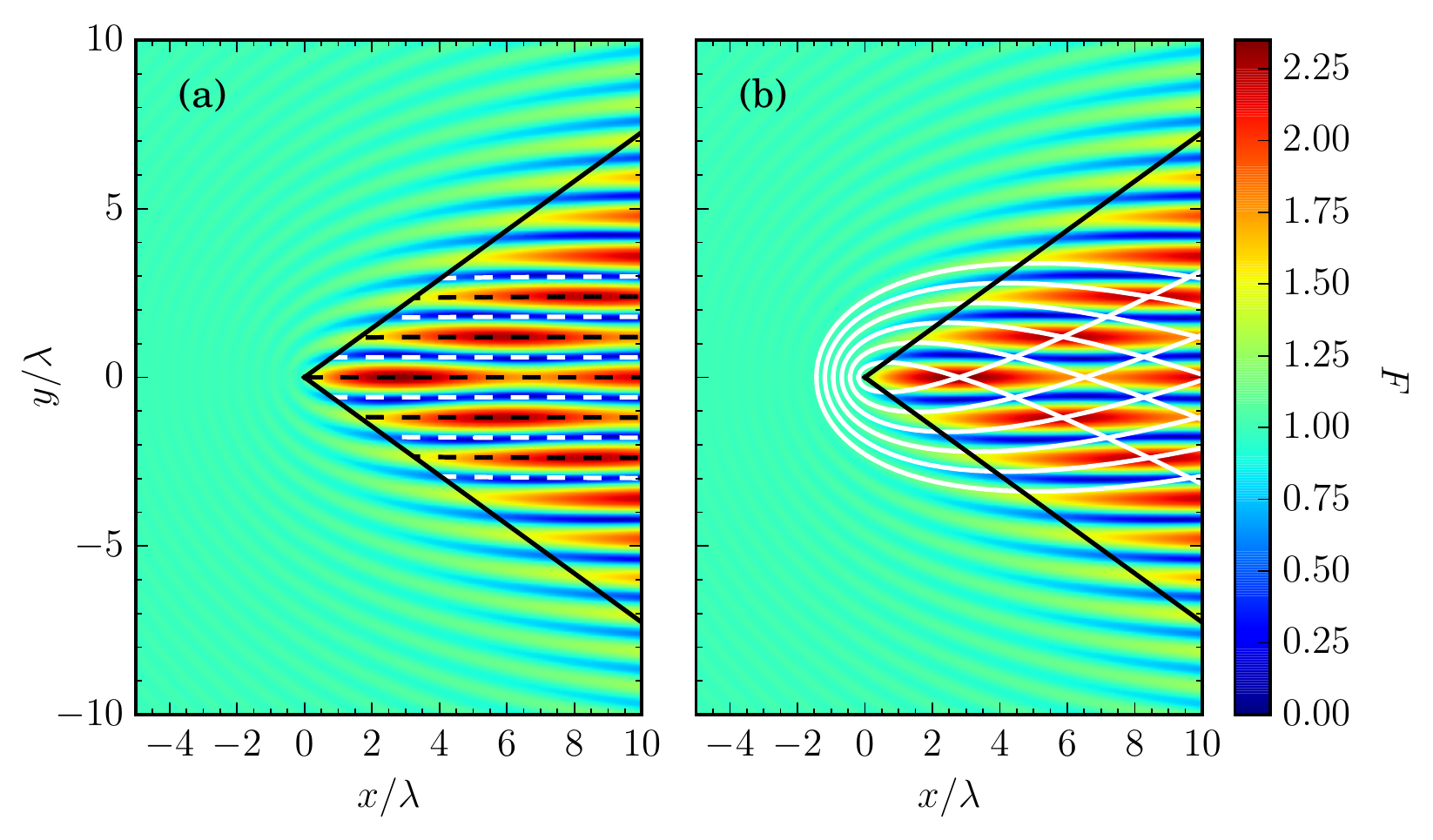}
\caption{Wave pattern for plane-wave propagation in conical space with
$\Delta=\pi/6$:
(a) white and black dashed lines correspond to the nodal and antinodal lines,
respectively. Intersection of white lines in (b) determine the diffraction
maxima.}
\label{fig:f-lines-pl}
\end{figure}

If the distance from the source to the string goes to infinity, $r_0 \to \infty$, 
one can neglect the effects of the wavefront curvature and consider
the incidence of a plane wave.
In this limit, the field at a point $(r,\varphi)$ is given by \cite{pla-string16}
\begin{equation}
E=\eee{\ii kr\cos(\Delta+\beta\varphi)}\mathcal{F}(u^+)+\eee{\ii
kr\cos(\Delta-\beta\varphi)}\mathcal{F}(u^-)
\label{eq:u-pl}
\end{equation}
with $u^\pm=\sqrt{2kr}\sin[(\Delta\pm\beta\varphi)/2]$. 
We did not carry out any numerical simulations for this limiting case, however,
it would be interesting to compare the analytical results \eref{eq:u-pl}
with the similar results for a finite-distant source.
For a plane wave, the nodal and antinodal lines should be straight lines in
$(r,\theta)$ space, parallel to the line of sight
\begin{equation}
2 r \sin\theta \,\sin\Delta = \frac{\lambda}{2} \, q
\label{eq:go-interf2}
\end{equation}
with $q$ being an integer. 
After the angular transformation $\theta=\beta\varphi$, those lines will still
be almost parallel to the $x$-axis, as long as $\sin(\beta\varphi) \approx
\beta\varphi$ for small arguments.
This is seen in figure \ref{fig:f-lines-pl}(a) where the spatial distribution of
the amplification factor $F$ (equal to $|E|^2$ in this case) is presented for
$\Delta=\pi/6$.
The typical spacing between the fringes along the $y$-axis, as follows from
equation \eref{eq:go-interf2}, is approximately $\delta
y\approx\lambda/(2\beta\sin\Delta)$, and it is independent of the distance from
the string.
For plane waves, the hyperbolic lines of constant phase between the GO and D
waves become parabolic, $r[1-\cos(\Delta\pm\theta)] = const$, and the intensity
maxima may be found at the intersection points of the curves
\begin{equation}\eqalign{
r [1 - \cos(\Delta + \beta\varphi)]  =  \frac{\lambda}{2} \, \left(n+\frac{3}{4} \right), \\
r [1 - \cos(\Delta - \beta\varphi)]   =  \frac{\lambda}{2} \,
\left(m+\frac{3}{4} \right)}
\label{eq:max-min3}
\end{equation}
with $m,n$ being non-negative integers, as it is depicted in figure
\ref{fig:f-lines-pl}(b).

Given an excellent correspondence with the numerical results, one can use
equations \eref{eq:max-min3} to predict the location of some typical maxima.
For instance, the position of the highest diffraction maximum can be estimated
by substituting $m=n=0$ and $\varphi=0$. For $\Delta=\pi/6$, this gives
$x\approx 2.8\lambda$.
The next-order maxima ($n=2$ and $m=0$ and vice versa) are at $x\approx
5.9\lambda$ and $y\approx \pm 1.2\lambda$, while the fringe separation for this
$\Delta$ should be $\delta y \approx 1.2\lambda$.
All these values are seen to be in agreement with figure \ref{fig:f-lines-pl}.


\section{Concluding remarks}

Transformation optics is a powerful tool which allows one to design an
artificial meta\-material medium in which the light behaves in a similar way as
if it were in the curved space.
In this paper, we have analyzed the medium properties and the propagation of
electromagnetic waves through the effective media which mimic the conical
topology of a static cosmic string - a 1D topological defect.
A conical spacetime is an example of a geometry which has a singular
$\delta$-like curvature at the string core.
This singularity has important physical consequences similar to those which
arise in the Aharonov-Bohm setting.
Photons or particles constrained to move in a region where the Riemann tensor
vanishes may nonetheless exhibit physical effects arising from non-zero
curvature confined to the string core, i.e., a region from which they are
excluded. This is a gravitational analogue of the Aharonov-Bohm  effect for a
static mass distribution \cite{dowker67,kraus68,ford-vilenkin81}.
The physical manifestations of this effect can be summarized as follows:
\begin{enumerate}
 \item In the geometrical-optics limit, the light beams or photons propagating
on different sides of the string can intersect, and their relative deflection
cannot, in general, be transformed away. This is analogous to the
electromagnetic Aharonov-Bohm effect, where the vector potential can be
transformed to zero only in regions containing no closed loops enclosing the
solenoid \cite{dowker67,kraus68,ford-vilenkin81}.
 \item The deflection angle $\sim \Delta$ is independent of the impact
parameter, but depends on the strength of the singularity, i.e., on the mass of
the string \cite{vilenkin81}.
 \item The beams intersection gives rise to a double imaging effect.
 \item In the double-imaging region, the interference pattern should appear with
the spacing between the fringes asymptotically $\sim \lambda/(2\beta\sin\Delta)$, which is independent of the distance. 
 \item The diffraction effects are manifested in the intensity modulation along
the antinodal lines and in the wave pattern beyond the geometrical-optics
domain.
 \item The Fresnel number $N_F \sim r \Delta^2/\lambda$ depends on the radial
coordinate, hence, the Fresnel observation zones can be introduced associated
with the intensity amplification caused by diffraction \cite{pla-fresnel17}.
\end{enumerate}
In the present paper, we are mainly concerned with the wave effects (iv) and (v).
We carry out the full-wave numerical simulation of the metamaterial medium which
mimics the cosmic-string topology.
The obtained results are interpreted by use of the asymptotic theories of
diffraction which, thus far, have been essentially applied to obstacles with
clearly defined boundaries (a half-plane, a slit, etc.\cite{borovikov}). Here,
we apply them to a topological defect and obtain the excellent agreement with the numerical
results.
We notice the advantage of the asymptotic theories. The existing solutions for
similar problems in the form of integral representation \cite{linet86} or
infinite series \cite{suyama06} exclude rigorous analysis necessary for
practical applications. 
The infinite series solution, for instance, is poorly convergent for the
distances greater than about a wavelength. These limitations are overcome by the
GTD and UAT we have used in this work, since they capture the main physics and a
few terms are only needed to achieve the required accuracy.

At first sight, the emergence of two image sources in conical space followed by their mutual interference, may look like a familiar Young's double-slit interference experiment. Our results, however, indicate that the two models are conceptually different. 
The Young's double-slit setting has a characteristic length -- the distance between the slits $d$, that implies that the interference fringe scales as $\sim \lambda\, r/d$ \cite{born-wolf-03}. On the contrary, for the conical space, there is no characteristic length but the deficit angle $\Delta$. 
As a result, the interference and diffraction effects manifest differently from those in the Young's experiment (see items (iv)--(vi) above). 
All of these peculiar features open up new opportunities for applications in
photonics and other fields. For instance, such metamaterial media with conical singularities can act as omnidirectional beam steering devices \cite{zhang15}, beam splitters \cite{zhang15,chang15,ahn15}, diffraction-control elements \cite{shi15}, etc.
Our results can also potentially be applied to light \cite{fumeron15,carvalho07} 
or sound propagation \cite{pereira13,fumeron17} 
near a linear topological defect in nematic liquid crystals. 
It would also be interesting to extend our approach to other geometries or types of
topological defects.


\ack
IFN acknowledges  financial  support  from  Universitat  de  Barcelona  under
the APIF scholarship.

\end{document}